\documentclass[preprint,authoryear,12pt]{elsarticle}

\usepackage{amssymb}

\usepackage[nodots]{numcompress}

\journal{New Astronomy}

\usepackage{epsfig}

\def\ut#1{\mathop{\vtop{\ialign{##\crcr
     $\hfil\displaystyle{#1}\hfil$\crcr\noalign
     {\kern1pt\nointerlineskip}\hbox{$\hfil\sim\hfil$}\crcr
     \noalign{\kern1pt}}}}}

\def\undersymbol#1#2{\mathop{\vtop{\ialign{##\crcr
     $\hfil\displaystyle{#2}\hfil$\crcr\noalign
     {\kern1pt\nointerlineskip}\hbox{$\hfil#1\hfil$}\crcr
     \noalign{\kern1pt}}}}}
\def\arcsec{^{\prime\prime}}
\def\arcmin{^{\prime}}
\def\degr{^0}
\def\hour{^{\rm h}}
\def\minute{^{\rm m}}
\def\second{^{\rm s}}

\begin{document}

\begin{frontmatter}



\title{The {\it XMM}-Newton Slew view of IGRJ17361-4441: a transient in the globular cluster NGC 6388}


\author[ad1,ad2]{A.A.Nucita}
\author[ad1,ad2]{F. De Paolis}
\author[ad3]{R. Saxton}
\author[ad4]{A.M. Read}
\address[ad1]{Department of Physics, University of Salento, Via per Arnesano, CP 193, I-73100, Lecce, Italy }
\address[ad2]{INFN, Sez. di Lecce, via Per Arnesano, CP 193, I-73100, Lecce, Italy }
\address[ad3]{European Space Astronomy Centre, SRE-O, P.O.~Box 78, 28691, Villanueva de la Ca\~nada (Madrid), Spain}
\address[ad4]{ Department of Physics and Astronomy, Leicester University, Leicester LE1 7RH, U.K}

\begin{abstract}
IGRJ17361-4441 is a hard transient recently observed by the $INTEGRAL$ satellite. The source, close to the center of gravity of the 
    globular cluster NGC 6388, quickly became the target of follow-up observations conducted by the Chandra, Swift/XRT and RXTE observatories. 
    Here, we concentrate in particular on a set of observations conducted by the {\it XMM}-Newton satellite during two slews, in order to get the spectral information of the source and
    search for spectral variations.
  The spectral parameters determined by the recent {\it XMM}-Newton slew observations were compared to the previously known results. 
  {The maximum unabsorbed $X$-ray flux in the 0.5-10 keV band as detected by the $XMM$-Newton slew observations is $\simeq 4.5\times 10^{-11}$ 
   erg cm$^{-2}$ s$^{-1}$, i.e. 
   consistent with that observed by the Swift/XRT satellite 15 days earlier. The spectrum seems to be marginally consistent 
   ($\Gamma\simeq 0.93-1.63$) with that derived from the previous high energy observations.}

\end{abstract}


\end{frontmatter}

\section{Introduction}
\label{s:introduction}

IGRJ17361-4441 is a hard X-ray transient source first observed by \citet{gibaud2011} 
with the IBIS/ISGRI telescope \citep{ubertini2003} onboard the {\it INTEGRAL} satellite \citep{winkler2003}
and was quickly recognized to be hosted in the galactic globular cluster NGC 6388 \citep{ferrigno2011}.

The location of the transient (close to the globular cluster gravitational center, but see later) 
is of great importance since NGC 6388, among all the globular clusters in our Galaxy, 
is one of the best candidates \citep{baumgardt2005} to host an intermediate mass black hole (hereafter IMBH). In particular,
by using  high resolution optical observations, \citet{lanzoni2007} estimated the mass of the IMBH to be $\simeq 5700$ 
M$_{\odot}$. 
It would be natural for such an IMBH to emit significant radiation 
in the X-ray band due to the likely accretion of matter from its surroundings. 
{In the context of the earliest observations of globular clusters, 
\citet{bahcall1975} and \citet{silk1976} were the first to suggest that the X-ray
emission detected towards these clusters was due to IMBHs (in the mass range $20$ M$_{\odot}$ -- $10^6$M$_{\odot}$) 
accreting from the intracluster medium. This issue was considered more recently by \citet{grindlay2001} who provided the census of the compact 
object and binary population in the globular cluster 47 Tuc and obtained an upper limit to the central IMBH of a few hundred solar masses.} 

Initial {\it XMM}-Newton and Chandra observations in this direction (\citealt{nucita2008} and \citealt{cseh2010}) showed that 
the core of NGC 6388 hosts several {\it X}-ray sources. 
{Based on the correlation between the X-ray and radio flux from black holes (\citet{merloni2003}); \citet{maccarone2004} 
was the first to point out that the search for radio emission from faint black holes is useful to test the IMBH hypothesis in globular clusters and dwarf
spheroidal galaxies (\citealt{maccarone2005}). \citet{cseh2010} 
observed the central region of NGC 6388 in the radio band using the Australia Telescope Compact Array (ATCA) to search for radio signatures of the IMBH.
The radio observation resulted in an upper limit of the IMBH mass of $\simeq 1500$ M$_{\odot}$.}

The discovery of a transient source close to the NGC 6388 gravitational center could be related to 
the turning on of the putative globular cluster IMBH. 
However, as will become clear in the subsequent sections,
the nature and spectral properties of the transient IGRJ17361-4441 are difficult to reconcile with the IMBH picture 
and rather favour an interpretation as a high mass X-ray binary (HMXB) or
a low mass X-ray binary (LMXB). 
Several observational campaigns (in the {\it X}-rays as well as in the 
radio band) were organized in order to pinpoint IGRJ17361-4441 and draw firm conclusions
on the NGC 6388 IMBH paradigm.

In this paper we briefly discuss (see section \ref{s:previousObs}) the past $X$-ray observations of the NGC 6388 globular cluster (see \citealt{nucita2008}, and 
\citealt{cseh2010}) and the discovery of 
the hard transient IGRJ17361-4441 by  {\it INTEGRAL} (\citealt{gibaud2011}) as well as the 
follow-up observations conducted by Chandra (\citealt{pooley2011}), 
Swift/XRT and RXTE observatories (\citealt{ferrigno2011} and \citealt{bozzo2011}).
Then we concentrate (see section \ref{s:xmmObs})
on the analsyis of two {\it XMM}-Newton slew observations of NGC 6388 conducted 15 days after the {\it INTEGRAL} 
discovery of the source. The two slew observations had $\simeq 7.6$ seconds and $\simeq 7.7$ seconds on source exposure time. Finally our conlusions are presented in 
section \ref{s:conclusion}.

\section{Previous observations of NGC 6388}
\label{s:previousObs}
\subsection{{\it XMM}-Newton and Chandra observations of NGC 6388}

By studying a combination of high resolution (HST ACS-HRC, ACS-WFC, and WFPC2) and wide field (ESO-WFI) observations
of the globular cluster NGC 6388, \citet{lanzoni2007} claimed the existence of a central IMBH. Such a compact object of mass
$\simeq 5700$ M$_{\odot}$ should reside in the globular cluster center of gravity\footnote{Note however that, as first pointed out by \citet{bahcall1976},
a black hole  in a stellar cluster will experience a Brownian motion due to gravitational interactions with the surrounding objects. 
Thus, the black hole is not necessarily  at the dynamical center of the host cluster, but may move with mean square velocity given by
(\citealt{merritt2007})
\begin{equation}
\frac{1}{2}M<v_{rms}^2>\simeq \frac{3}{2}m\sigma^2
\label{vrms}
\end{equation}
where $M$, $m$ and $\sigma$ represent the black hole mass, the perturber average mass and the stellar velocity 
dispersion within $\sim 0.6 r_i$, respectively. Here $r_i$ is the influence radius of the black hole (for details see \citealt{merritt2007} 
and references therein).} localized at the coordinates (J2000) 
$RA=17\hour~36\minute~17.23\second$, $Dec=-44\degr~44\arcmin~7.1\arcsec$. An uncertainty of $0.3\arcsec$ is associated with both coordinates.

\citet{nucita2008} suggested that this IMBH should emit radiation in the $X$-ray band due to accretion from 
the surrounding matter. A $48$ ks {\it XMM}-Newton observation was made on 21 March 2003. It resulted in a spectrum which was 
well fit by an absorbed power-law model. The resulting best fit 
parameters were $N_H=(2.7\pm0.3)\times 10 ^{21}$ cm$^{-2}$ for the hydrogen column density and $\Gamma=2.4\pm0.1$ for the power law index.
The unabsorbed flux in the $0.5-7$ keV band was $F_{0.5-7}=(4.0\pm0.2)\times 10^{-13}$ erg cm$^{-2}$ s$^{-1}$ which, 
for a distance of $13.2$ kpc corresponds to a luminosity
of $L_{0.5-7}\simeq (7.2\pm 0.4)\times 10^{33}$ erg s$^{-1}$. Note that the hydrogen column density is consistent with the average one 
found in the direction of the target (\citealt{dickey1990}). 

The Chandra satellite, with a much better angular resolution than that of $XMM$-Newton, observed towards NGC 6388 for $\simeq 45$ ks on 
21 April 2005 (id 5505). \citet{nucita2008} identified 16 discrete 
sources within the half mass radius ($\simeq 40\arcsec$, see \citealt{lanzoni2007}) of the cluster. The $3$ sources close to the gravitational center
were not spatially resolved by the authors, so that they were considered virtually as a single source (labeled as $\#14^*$). 
The unabsorbed flux in the 0.5-7 keV band 
of the $\#14^*$ is $F_{0.5-7}\simeq 1.7\times 10^{-13}$ erg cm$^{-2}$ s$^{-1}$, corresponding to a luminosity of 
$L_{0.5-7}\simeq 3\times 10^{33}$ erg s$^{-1}$.

A more detailed analysis on the same Chandra data set was conducted by \citet{cseh2010}. After removing the pixel randomization, these authors were able
to spatially resolve the source $\#14^*$ into three separate sources labeled as $\#12$, $\#7$ and $\#3$. In particular, the source $\#12$, which is consistent 
with the position of the center of gravity of NGC 6388, is characterized by an unabsorbed flux of $F_{0.3-8}\simeq 4.0\times 10^{-14}$ erg cm$^{-2}$ s$^{-1}$
corresponding to an intrinsic luminosiy of $L_{0.5-7}\simeq 8.3\times 10^{32}$ erg s$^{-1}$. 

{\citet{cseh2010} searched for a radio counterpart of the putative IMBH in NGC 6388 
using the ATCA facility}. Unfortunately, this search only resulted in an upper limit to the 
radio flux at 5 GHz of $\simeq 81$ $\mu$Jy/beam. Therefore, it was only possible to determine an upper limit to the IMBH radio 
luminosity of $L_R < 8.4\times 10^{28}$ erg s$^{-1}$. 

Based on the fundamental plane of black hole accretion (\citealt{merloni2003} and \citealt{kording2006}) and using the observed $X$-ray and radio luminosities, 
it was then possible to put a 3$\sigma$ upper limit of  $\simeq 1500$ M$_{\odot}$ on the mass of the IMBH in NGC 6388 (\citealt{cseh2010}). 
The estimated mass value has to be treated with caution for two reasons: {\it i}) the identification of the $X$-ray counterpart of such a black hole 
is not trivial since several sources are close to the NGC 6388 center of gravity. If none of them are associated with the IMBH, 
then one can not use the fundamental plane relation to get an estimate of the mass; {\it ii}) the fundamental plane relation 
(as derived by \citealt{merloni2003} and \citealt{kording2006}) is not tested for black hole masses in the range of interest 
for IMBHs, i.e. $10^3$ M$_{\odot}$ -- $10^4$ M$_{\odot}$. {Note however that \citet{maccarone2005} and \citet{maccarone2008} showed that the non-detection of a
radio source, in combination with the estimate of the globular cluster ISM density\footnote{The amount of gas contained in globular clusters is an issue of debate.
The intracluster medium density can be estimated by using the dispersion measures of the pulsars observed within the cluster 
(see e.g. \citealt{freire2001, freire2003}) or inferred by the empirical knowledge about the stellar mass loss (\citealt{pfahl}).}
and the expected value of the accretion rate, can be used to get information 
(at least as an order of magnitude estimate) of the IMBH mass.}

\subsection{{\it INTEGRAL} discovery of IGRJ17361-4441 and subsequent $X$-ray follow-up observations}

On 11 August 2011, \citet{gibaud2011} reported the discovery of a new hard $X$-ray transient (IGRJ17361-4441) by the IBIS/ISGRI 
telescope \citep{ubertini2003} onboard the {\it INTEGRAL} satellite \citep{winkler2003}. The spectrum of the source, 
associated with the globular cluster NGC 6388, was described by a power law with photon index $\Gamma=2.6^{+1.0}_{-0.7}$ and characterized by a flux 
in the 20-100 keV of $F_{20-100}\simeq 9.7\times 10^{-11}$ erg cm$^{-2}$ s$^{-1}$.

Since this newly discovered transient is possibly associated with the IMBH in NGC 6388, IGRJ17361-4441 
became the target of several $X$-ray follow-up observations aimed at obtaining accurate position and flux measurements
to test if this association is correct.

\citet{weinands2011} (but see also \citealt{ferrigno2011}) reported a Swift/XRT observation (1.9 Ks on 16 August 2011) 
in which the astrometrically corrected position of 
IGRJ17361-4441 was $RA=17\hour~36\minute~17.5\second$, $Dec=-44\degr~44\arcmin~7.1\arcsec$.
A more detailed analysis on the XRT data conducted by \citet{bozzo2011} determined the new transient position 
to be $RA=17\hour~36\minute~17.27\second$, $Dec=-44\degr~44\arcmin~7.0\arcsec$ with an associated error (on both coordinates) of $1.9\arcsec$. 
Thus, the distance\footnote{We note that the distances between the sources and the center of gravity of the globular cluser NGC 6388 are calculated by using the well
know Haversine formula. Distance uncertainties are calculated by corrctly propagating the errors on both $\alpha$ and $\delta$ coordinates.} 
of the transient from the center of gravity of NGC 6388 is $0.4\arcsec \pm 1.4\arcsec$. Hence, the source position is consistent 
(according to the Swift/XRT data) with the center of gravity of the cluster and (possibly) associated with the IMBH. Note also that 
a 2.5 Ks Chandra observation was made on 29 August 2011 (\citealt{weinands2011}) in order 
to improve the accuracy of the location of the transient.  
IGRJ17361-4441 is located at the coordinates $RA=17\hour~36\minute~17.418\second$, $Dec=-44\degr~44\arcmin~5.98\arcsec$ 
({the nominal Chandra positional accuracy is $0.6\arcsec$ on both coordinates}). In this case,
the estimated distance of the transient to the cluster center of gravity is $2.3\arcsec\pm0.5\arcsec$. 

Based on the Swift/XRT astrometry only, one could conclude that the position of IGRJ17361-4441 is formally consistent with the center of gravity of NGC 6388 and 
possibly related to the putative IMBH in the globular cluster. An updated radio observation conducted at ATCA by \citet{bozzo2011} put a more stringent upper 
limit to the radio luminosity of $L_R< 5\times 10 ^{28}$ erg s$^{-1}$ so that, following the same procedure as in \citealt{cseh2010} and the 2005 Chandra $X$-ray 
flux estimate, the new IMBH upper limit turns out to be $\simeq 600$ M$_{\odot}$ (\citealt{bozzo2011}). 

However, a caveat on this conclusion is necessary. The new Chandra refined source coordinates (even if formally consistent with the source position 
determined by Swift/XRT) indicate that the transient could be a new $X$-ray source (see later) not associated with the IMBH. In this case, and for the reasons explained above, 
one should not use the black hole fundamental plane relation in order to estimate the IMBH mass. 
If one believes that the transient is associated with the NGC 6388 center of gravity, then it should also be noted that 
at least three sources (those labeled as $\#12$, $\#7$ and $\#3$ in \citealt{cseh2010}) are 
within the error box of Swift/XRT. In particular, sources $\#12$ and $\#7$ have fluxes
$\simeq 4.0\times 10^{-14}$ erg cm$^{-2}$ s$^{-1}$ and $\simeq 6.9\times 10^{-14}$ erg cm$^{-2}$ s$^{-1}$, respectively. 
If source $7$ is associated with the IMBH, the $X$-ray and radio observations together with the fundamental plane relation give 
an upper limit of $\simeq 1200$ M$_{\odot}$.

The Swift/XRT spectrum of the transient source was fitted (\citealt{bozzo2011}) with an absorbed power law with photon index $\Gamma \sim 0.5-0.9$ and hydrogen column density 
$N_H\simeq (0.5-0.9)\times 10 ^{22}$ cm$^{-2}$, i.e. consistent with that derived from the previous $XMM$-Newton data. 

The flux in the $1.0-10$ keV band is 
$F_{1-10}= (4.5-4.8)\times 10^{-11}$ erg cm$^{-2}$ s$^{-1}$, a factor $100$ more luminous 
than the source $\#12$ possibly associated with the NGC 6388 IMBH (\citealt{cseh2010}). When the broad band spectrum (obtained by using Swift/XRT and {\it INTEGRAL/ISGRI}) was analysed 
(and fit with a broken power law), 
\citet{bozzo2011} obtained a flux of $F_{1-10}= (4.6^{+0.1}_{-0.5})\times 10^{-11}$ erg cm$^{-2}$ s$^{-1}$ in the $1.0-10$ kev band and 
a flux of $F_{20-100}= (7.8^{+0.8}_{-3.8})\times 10^{-11}$ erg cm$^{-2}$ s$^{-1}$ in the $20-100$ kev band. 

These results are consistent with those obtained by using the XRTE/PCA follow-up observation made on 17 August 2011. In particular, it was found that   
$F_{3-15}= (6.7^{+0.1}_{-3.4})\times 10^{-11}$ erg cm$^{-2}$ s$^{-1}$ in the $3-15$ keV band (\citealt{bozzo2011}).

\section{The {\it XMM}-Newton slew observations}
\label{s:xmmObs}
The large collecting area of the nested mirrors together with the high quantum efficiency of the EPIC/PN camera 
make the $XMM$-Newton satellite the most sensitive $X$-ray observatory available at present (\citealt{jansen2001}). 
The $XMM$-Newton satellite was recognized to be a good instrument to collect data during slewing manoeuvres and is performing an
$X$-ray survey of the sky (\citealt{saxton2008}). Note that the $XMM$-Newton slew observations are moderately deep with a detection limit 
of $1.2\times 10 ^{-12}$ erg cm$^{-2}$ s$^{-1}$ in the $0.2-12.0$ keV band. 

\begin{figure}[htbp]
 \vspace{6.5cm} \includegraphics{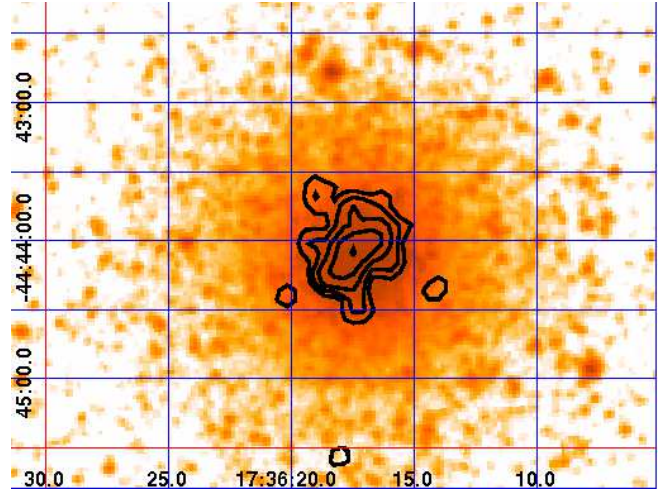}
 \caption{Contours (increasing by factors of 2) of lightly-smoothed 0.2-12 keV XMM-Newton slew
emission (EPIC-pn camera, the two 01/09/11 observations combined),
superimposed on a SAO-DSS image of NGC6388.} 
 \label{f0}
 \end{figure}

Due to the scheduled observation program, the $XMM$-Newton satellite twice observed the region of the sky around NGC 6388. 
The observations were taken on September 1$^{st}$ 2011 at 13:10:34 (herafter S1 observation) and 19:00:17 (UT) (herafter S2 observation), i.e. 
15 days after the first Swift/XRT follow-up observation of IGRJ17361-4441. The transient
source was then observed serendipitously\footnote{The exposure times were estimated by calculating the distance travelled 
by the source on the detector with the typical {\it XMM}-Newton slew speed of $90$ deg h$^{-1}$. The resulting exposure times were also corrected for chip gaps
 and bad pixels (for details see \citealt{read2008}).} for
$\simeq 7.6$ s and $\simeq 7.7$ s in the two slew observations with the EPIC/PN instrument (see Fig. \ref{f0}).
 \begin{figure}[htbp]
 \vspace{8.2cm} \includegraphics{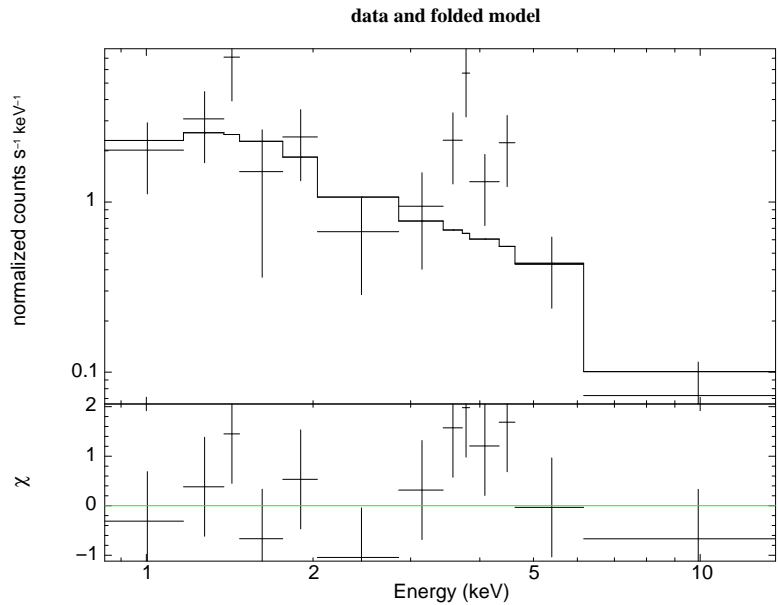}
 \caption{The $XMM$-Newton spectrum of IGRJ17361-4441 (data points) collected during the first slew observation on September 1$^{st}$ 2011. The solid line 
  represents the best fit model (see text for details).} 
 \label{f1}
 \end{figure}

 \begin{figure}[htbp]
 \vspace{8.2cm} \includegraphics{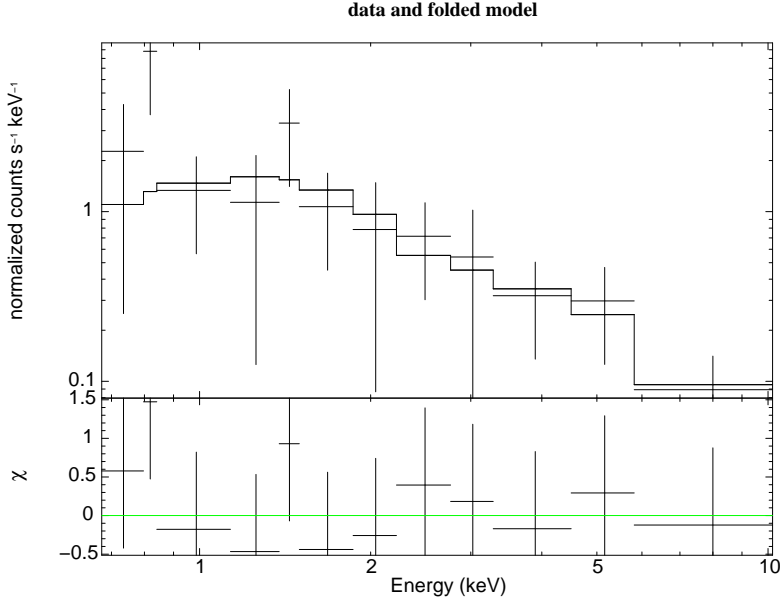}
 \caption{The same as in figure 1, but for the second $XMM$-Newton slew observation on September 1$^{st}$ 2011.} 
 \label{f2}
 \end{figure}

We decided to analyze the data sets separately in order to study a possible spectral variation on time-scales of hours. 

The source spectrum has been extracted from a circle of radius $60\arcsec$ about the source with the 
background being extracted from an annulus of inner radius $90\arcsec$ and outer radius $120\arcsec$ about the source.
The detector matrices are calculated taking into account
the transit of the source across the detector and using the method described in \citet{read2008}. 
Hence, the source and background spectra (as well as the response matrices) were imported in the XSPEC package 
(version 12.4.0) for the spectral analysis and fitting procedure.
The adopted model is an absorbed power law ({\it wabs*power}) with the hydrogen column density fixed to the average value 
found by ROSAT in the direction of the target, i.e. $2.5\times 10^{21}$ cm$^{-2}$. 
Note that this value is consistent with that derived by \citet{nucita2008} when analyzing the 2005 $XMM$-Newton observation of NGC 6388 and also similar
to the column density found by \citet{bozzo2011} (see also \citealt{ferrigno2011}) while studying the Swift/XRT follow-up observation of IGRJ17361-4441.
The adopted model has two free parameters; the photon index $\Gamma$ and the power law normalization $N$.

The fitting procedure to the S1 spectrum resulted in the best fit parameters ($\chi^2/\nu=1.4$ for 11 d.o.f.)  $\Gamma = 1.16\pm0.20$ and 
$N=(1.7\pm0.7)\times 10^{-3}$. 
The absorbed fluxes in the $0.5-2.0$ keV, $2.0-10$ keV, and $0.5-10.0$ keV bands are 
$F_{0.5-2}= (5.4^{+2.8}_{-3.2})\times 10^{-12}$ erg cm$^{-2}$ s$^{-1}$, $F_{2-10}= (3.6^{+1.4}_{-1.7})\times 10^{-11}$ erg cm$^{-2}$ s$^{-1}$, and 
$F_{0.5-10}= (4.1^{+1.6}_{-1.9})\times 10^{-11}$ erg cm$^{-2}$ s$^{-1}$, respectively.

Fitting the S2 spectrum with the same model gives the best 
fit parameters ($\chi^2/\nu=0.4$ for 10 d.o.f.)  $\Gamma = 1.28\pm0.35$ and $N=(1.2\pm0.6)\times 10^{-3}$. 
The absorbed fluxes in the same bands as above are  
$F_{0.5-2} = (3.4^{+2.3}_{-2.3}) \times 10^{-12}$ erg cm$^{-2}$ s$^{-1}$, $F_{2-10}= (1.9^{+1.1}_{-1.4})\times 10^{-12}$ erg cm$^{-2}$ s$^{-1}$, and 
$F_{0.5-10}= (2.1^{+1.2}_{-1.8})\times 10^{-11}$ erg cm$^{-2}$ s$^{-1}$, respectively.

The maximum unabsorbed flux of the source in the 0.5-10 keV band is $\simeq 4.5\times 10^{-11}$ erg cm$^{-2}$ s$^{-1}$ corresponding to an intrinsic luminosity 
of $\simeq 9.3\times 10^{35}$ erg s$^{-1}$. 

\begin{figure}[htbp]
 \vspace{7.cm} \includegraphics{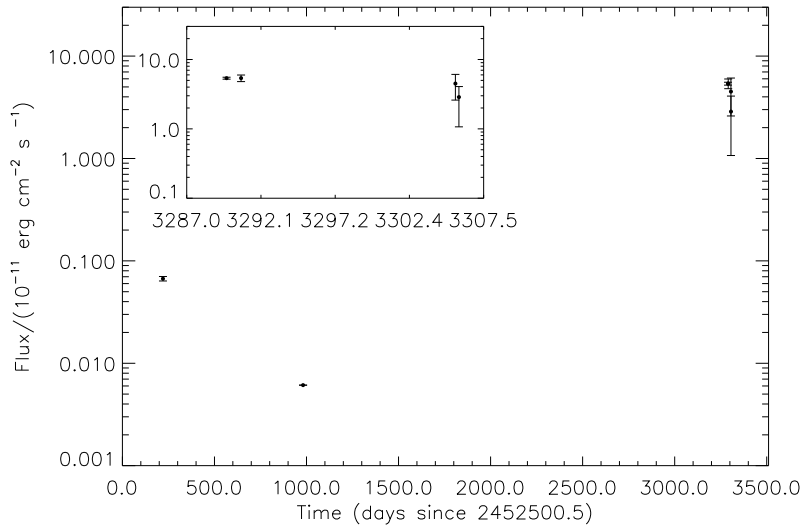}
 \caption{The data points from the left to the right correspond to the NGC 6388 flux in the $0.5-10$ keV band from 2003 to 2011. In the insert, 
we give the data points corresponding to the flare observed and monitored in 2011 only (see text for details).} 
 \label{f3}
 \end{figure}

In Figure \ref{f3}, the unabsorbed X-ray fluxes of NGC 6388 in the $0.5-10$ keV band (from 2003 to 2011) are shown. In the insert, we give
the data points corresponding to the IGRJ17361-4441 flare observed and monitored by several instruments (Swift/XRT, RXTE/PCA and $XMM$-Newton) in 2011 only. 
Note that for the $XMM$-Newton observation in 2003, the Chandra observation in 2005, and the Swift and RXTE in 2011 the $0.5-10$ keV band 
fluxes were obtained by extrapolating (to this energy band) the best fit models available 
in the literature (see e.g. \citealt{bozzo2011}).
 
What can be easily observed from the results reported here, is that the $X$-ray flux in the 0.5-10 keV band 
as detected by the $XMM$-Newton slew observations, i.e. $\simeq 4.5\times 10^{-11}$ erg cm$^{-2}$ s$^{-1}$, is consistent with that
observed by Swift/XRT 15 days earlier. {Note also that the spectrum power-law seems to be marginally consistent  ($\Gamma\simeq 0.93-1.63$) with that
derived from the previous high energy observations ($\Gamma\simeq 0.5-0.9$, see \citealt{bozzo2011}), a point that could help in a classification of 
the transient nature (see the subsequent discussion).}

\section{Results and discussion}
\label{s:conclusion}
IGRJ17361-4441 is a hard transient recently observed by the $INTEGRAL$ satellite. $X$-ray follow-up observations have shown that 
the source is within the globular cluster NGC 6388. 
Based only on the astrometry of the Swift/XRT satellite, the transient position is consistent with the center of gravity of the globular cluster and  
this opens the possibility that IGRJ17361-4441 is associated with an IMBH which is turning on.  

However, if one believes that the transient is associated with the IMBH in NGC 6388, then it should be noted that 
at least three $X$-ray sources (those labeled as $\#12$, $\#7$ and $\#3$ in \citealt{cseh2010}) are within the error box of Swift/XRT. 
In particular, the sources $\#12$ and $\#7$ have fluxes which differ at least by a factor $2$ between them. 
If source $\#7$ is associated with the IMBH, then the observed Chandra $X$-ray flux (see \citealt{nucita2008} and \citealt{cseh2010}) and 
the updated radio observation of \citet{bozzo2011} together with the fundamental plane relation give  an upper limit of $\simeq 1200$ M$_{\odot}$.

In the IMBH hypothesis, the intrinsic luminosity of the source as determined by using the $XMM$-Newton slew data 
(i.e. $\simeq 9.3\times 10^{35}$ erg s$^{-1}$) should be
compared with that derived by using the 2005 Chandra data (and in particular for the source $\#12$ in \citealt{cseh2010}) 
when the putative IMBH was in quiescent state. In this case, one finds that the source luminosity increased by at least a factor $\simeq 1000$. 
Moreover, the spectrum seems to follow a power law with photon index $\Gamma\simeq 0.96-1.63$.

Nevertheless, the refined source position given by the Chandra satellite 
(even if still in agreement with the Swift/XRT result) 
argues against the IMBH hypothesis in favor of a newly discovered source. 
In this case, the $XMM$-Newton intrinsic source luminosity should be compared with the upper limit for the 
quiescent state of the source. \citet{pooley2011}, based on the non-detection of the source in the 2005 Chandra observation, estimated 
this limit to be $\simeq 10^{31} $erg s$^{-1}$. Thus, in this case the transient source has increased its luminosity by a factor 
close to $10^5$.

Two more possibilities for the nature of the transient source are that it is either a HMXB or a LMXB. 
{The first possibility is actually unlikely since these systems involve companion stars with mass 
larger than $\simeq 10$ M$_{\odot}$ (\citealt{lewin2006}), i.e. O/B stars which are not expected to exist in globular clusters. 
Note also that NGC 6388 was extensively observed by the HST instruments (see e.g. \citealt{hst}) and the collected data
did not show the presence of any O or B star in the globular cluster.

Hence, among the IMBH alternatives, the LMXB option is the most favorable. 
This is supported by the $X$-ray luminosity ($\simeq 9.3\times 10^{35}$ erg s$^{-1}$) and by the soft spectrum 
($\Gamma\simeq 0.93-1.63$) observed in the
$XMM$-Newton slew observation which seems to be consistent with the typical characteristics of the LMXB class of objects. 
A long $X$-ray observation (sufficient to allow a detailed timing and spectral analysis) may 
help in understanding the physics underlying this transient source.}

%


\end{document}